\documentclass[aps,prl,twocolumn,preprintnumbers]{revtex4-2}
\usepackage{amsmath, mathrsfs, amssymb,amsfonts,amsthm,graphicx, epsf, dcolumn, yfonts, subfigure}
\usepackage[hyperfootnotes=true]{hyperref}
\usepackage{xcolor}
\usepackage{slashed,comment}
\usepackage{setspace}
\usepackage{cancel}
\usepackage{wasysym}
\usepackage{float}
\usepackage[normalem] {ulem}
\pdfoutput=1
 \newcommand{\be}{\begin{equation}}
 \newcommand{\ee}{\end{equation}}
 \newcommand{\bea}{\begin{eqnarray}}
 \newcommand{\eea}{\end{eqnarray}}

\newcommand{\exd}{\mathrm{d}}

\newcommand{\beq}{\begin{equation}}
\newcommand{\eeq}{\end{equation}}

\def\r{\rho}

\renewcommand*{\thefootnote}{\fnsymbol{footnote}}

\begin{document}

\preprint{\texttt{IFT-UAM/CSIC-25-3}}

\title{Quantum induced superradiance}
\author{Casey Cartwright,$^{1}$ Umut G\"ursoy,$^{1}$ Juan F. Pedraza$^{2}$ and Andrew Svesko$^{3}$}
\affiliation{\vspace{1mm}
$^1$Institute for Theoretical Physics, Utrecht University, 3584 CC Utrecht, The Netherlands\\
$^2$Instituto de F\'isica Te\'orica UAM/CSIC, Cantoblanco, 28049 Madrid, Spain\\
$^3$Department of Mathematics, King’s College London, Strand, London, WC2R 2LS, UK}

\begin{abstract}\vspace{-2mm}

 \noindent Superradiance, the phenomenon enabling energy extraction through radiation amplification, is not universal to all black holes. We show that semi-classical backreaction can induce superradiance, even when absent at the classical level. Specifically, we compute the quasinormal modes of a massless scalar field probing a family of rotating `quantum' black holes in three-dimensional anti-de Sitter space, accounting for all orders of backreaction due to quantum conformal matter. A subset of these modes is identified as superradiant, leading to the formation of quantum black hole `bombs'. All such quantum black holes have curvature singularities shrouded by horizons. Thus, while backreaction enforces cosmic censorship, it also renders the black holes dynamically unstable. Further, we find all thermally unstable black holes  are dynamically unstable, though the converse does not hold generally. Our findings thus suggest a semiclassical version of the Gubser-Mitra conjecture on black hole stability. This motivates us to propose a stability criterion for quantum black holes.
\end{abstract}

\renewcommand*{\thefootnote}{\arabic{footnote}}
\setcounter{footnote}{0}

\maketitle

\noindent \textbf{Introduction.}
Like the Penrose process for extracting energy from a rotating black hole \cite{Penrose:1969pc}, superradiance occurs when a reflected wave has a larger amplitude than the incoming wave \cite{1971JETPL..14..180Z,1972JETP...35.1085Z,Misner:1972kx,Press:1972zz,Teukolsky:1974yv,Bekenstein:1998nt,Brito:2015oca}.
To wit, a wave of frequency $\omega$ and azimuthal number $m$ incident on a Kerr black hole of mass $M$ and spin $J$ yields the change $dM/dJ=\omega/m$. The first law of black hole mechanics then reads \cite{Bekenstein:1973mi}
\beq dM=\frac{\omega\kappa_{g}}{8\pi G}\frac{dA}{(\omega-m\Omega)}\;, \label{eq:firstlaw}\eeq
for black hole surface gravity $\kappa_{g}$, horizon area $A$, and angular velocity $\Omega$. Assuming the wave obeys the weak energy condition, such that the horizon area never decreases \cite{Hawking:1971tu}, amplification occurs when $\omega<m\Omega$,
yielding black hole energy extraction, $dM<0$.\footnote{Analogous arguments hold for superradiant amplification of charged waves scattered off charged black holes \cite{Denardo:1973pyo,Damour:1974qv,Gibbons:1975kk}, where $\omega<q\Phi$ for electric wave charge $q$ and black hole horizon electrostatic potential $\Phi$.}

Black hole superradiance predates evaporation via thermal radiation \cite{Hawking:1974sw} and serves as the foundation for the Blandford-Znajek mechanism \cite{Blandford:1977ds}, thought to drive jet formation in astrophysical black holes. Moreover, since repeated amplification can trigger a dynamical instability \cite{Teukolsky:1974yv,Damour:1976kh,Zouros:1979iw,Detweiler:1980uk}, superradiance is central to black hole stability and the ultimate fate of gravitational collapse.

Not all black holes exhibit superradiance, however, nor do all types of classical waves. Firstly, superradiance occurs when scattering bosonic fields off of Kerr-Newman (KN) black holes, but not fermionic fields \cite{Unruh:1973bda}. Further, large four-dimensional KN black holes with anti-de Sitter (AdS) asymptotics have no superradiance if the bosonic fields obey reflective boundary conditions at the timelike boundary at spatial infinity \cite{Hawking:1999dp}; superradiance can be restored if the fields obey transparent boundary conditions \cite{Winstanley:2001nx}. Another notable exception is the AdS$_{3}$ Ba\~nados-Teitelboim-Zanelli (BTZ) black hole \cite{Banados:1992wn,Banados:1992gq}, which does not exhibit superradiance for neutral scalar or fermionic fields obeying Dirichlet boundary conditions at spatial infinity \cite{Ortiz:2011wd,Ortiz:2018ddt}. Real scalars obeying generalized boundary conditions can lead to superradiance of the BTZ \cite{Iizuka:2015vsa,Dappiaggi:2017pbe,Ferreira:2017cta}, although not all superradiant modes are unstable.\footnote{The charged BTZ black hole can exhibit superradiance due to charged real scalar fields subject to mixed boundary conditions \cite{Konewko:2023gbu}.}

The aforementioned examples of black hole superradiance are wholly \emph{classical}. Both the black hole spacetime and incident wave are treated classically, adhering to general relativity. It is natural to wonder how semiclassical quantum effects modify the situation. Specifically, if there are scenarios where, classically, there is no superradiance, but quantum effects instigate its onset. Particularly challenging is the influence of semiclassical backreaction due to quantum fields in an otherwise classical spacetime. A complete treatment requires solving the semiclassical Einstein equations, a difficult open problem.

 Here we show quantum backreaction can induce black hole superradiance. Our study centers on an \emph{exact} black hole in semiclassical gravity, the `quantum' BTZ (qBTZ) black hole \cite{Emparan:2002px,Emparan:2020znc}.
 The qBTZ solution arises from braneworld holography~\cite{deHaro:2000wj}, a framework where a $(D-1)$-dimensional end-of-the-world brane is coupled to general relativity in a $D$-dimensional asymptotically AdS space, which has a dual description as a conformal field theory (CFT) living on the AdS boundary. A higher curvature gravity theory is induced on the brane coupled to a CFT with large central charge and an ultraviolet cutoff. In this formalism, AdS black holes that localize on branes may be understood as quantum-corrected black holes from the brane perspective, and can be exactly constructed to all orders of backreaction in three-dimensions \cite{Emparan:2002px,Panella:2024sor}.

\noindent \textbf{Quantum BTZ black holes.} We focus on the quantum BTZ (qBTZ) black hole \cite{Emparan:2020znc}. Its construction relies on an $\text{AdS}_{3}$ brane \cite{Karch:2000ct} intersecting the AdS$_{4}$ C-metric black hole horizon \cite{Emparan:1999wa,Emparan:1999fd}. Aided by holography, the geometry and thermodynamics of the qBTZ are known analytically and non-perturbatively in backreaction. Here we summarize the essentials of the solution. Further details can be found in the supplemental material.

In Boyer-Lindquist coordinates $(\bar{t},\bar{r},\bar{\phi})$, the neutral qBTZ black hole has line element \cite{Emparan:2020znc}
\beq
\hspace{-7mm}
\begin{split}
ds^2 &= -\left(\frac{\bar{r}^2}{\ell_3^2}-8\mathcal{G}_3 M -\frac{\ell \mu \eta^2}{r} \right) d\bar{t}^2 \\
&\hspace{-7mm} + \left(\frac{\bar{r}^2}{\ell_3^2}-8\mathcal{G}_3 M + \frac{(4 \mathcal{G}_3 J)^2}{\bar{r}^2}- \ell \mu (1-\tilde{a}^2)^2 \eta^4 \frac{r}{\bar{r}^2} \right)^{\hspace{-2mm}-1} \hspace{-1mm}d\bar{r}^2  \\
&\hspace{-7mm} + \left(\bar{r}^2 + \frac{\mu \ell\tilde{a}^2 \ell_3^2 \eta^2}{r} \right) d\bar{\phi}^2 - 8 \mathcal{G}_3 J \left( 1+ \frac{\ell}{x_1 r} \right) d\bar{\phi} d\bar{t}\;.
\end{split}
\label{eq:qbtzrotatmetmain}\eeq
Here, $r$ is a function of coordinate $\bar{r}$, which is kept for convenience, and $\mu$ is a mass parameter for the AdS$_{4}$ black hole.
Quantities $M$ and $J$ denote the mass and angular momentum of the black hole
\beq
\begin{split}
&8 \mathcal{G}_3 M = 4 \frac{- \kappa x_1^2 + \tilde{a}^2(4-\kappa x_1^2)}{(3-\kappa x_1^2 - \tilde{a}^2)^2}\;,\\
&4 \mathcal{G}_3 J=\frac{4 \ell_{3}\tilde{a}(1-\kappa x_1^2 + \tilde{a}^2)}{(3 - \kappa x_1 ^2 -\tilde{a}^2)^2}\;,
\end{split}
\label{eq:MJqbtz}\eeq
with $\eta\equiv 2x_{1}/(3-\kappa x_{1}^{2}-\tilde{a}^{2})$. Here $\ell$ is the UV cutoff length scale,
  $\mathcal{G}_3=G_{3}/\sqrt{1+(\ell/\ell_{3})^{2}}$ is the `renormalized' Newton's constant, and $\ell_{3}$ is the $\text{AdS}_{3}$ length.
Further, $\tilde{a}\equiv a x_{1}^{2}/\ell^{3}$, for rotation parameter $a$, $\kappa =\pm1,0$ corresponds to different brane slicings, and $x_{1}$ is a real parameter whose precise meaning is not relevant for us here.

The inner/outer event horizons $\bar{r}_{\pm}$ are generated by orbits of the Killing vector $\chi=\partial_{\bar{t}}+\Omega_{\pm}\partial_{\bar{\phi}}$,
where $\Omega_{\pm}$ denotes the horizon angular velocity relative to a \emph{non-rotating} frame at spatial infinity
\beq \Omega_{\pm}\equiv \frac{a}{r_{\pm}^{2}+a^{2}x_{1}^{2}}\left(1+\frac{r_{\pm}^{2}x_{1}^{2}}{\ell_{3}^{2}}\right)=\frac{d\bar{\phi}}{d\bar{t}}\;.\label{eq:Omnonrotinf}\eeq
The ergoregion corresponds to orbits of the Killing vector $\partial_{\bar{t}}$, where the $\bar{t}\bar{t}$-metric component of (\ref{eq:qbtzrotatmetmain}) vanishes.

The metric (\ref{eq:qbtzrotatmetmain}) is understood as a \emph{quantum} black hole as it is a solution to the induced semiclassical theory on the brane at all orders in quantum backreaction. The dimensionless parameter $\nu\equiv\ell/\ell_{3}$ controls the strength of backreaction, while the geometry (\ref{eq:qbtzrotatmetmain}) represents the quantum-corrected black hole due to the backreacting CFT$_{3}$. This is most easily seen in the static limit ($J=0$), where the metric takes the same form as the quantum-corrected BTZ solution due to perturbative backreaction of a conformally coupled scalar \cite{Steif:1993zv,Casals:2016ioo}. Notably, however, the metric (\ref{eq:qbtzrotatmetmain}) is valid for any strength of backreaction, i.e., any $\nu\geq0$. As $\nu \to 0$ with fixed $c_3$, backreaction vanishes, gravity weakens, and the metric resembles the classical BTZ black hole. Conversely, large-$\nu$ corresponds to strong backreaction; in this limit higher-derivative terms in the induced action become important.

Combined, $(a,\kappa,x_{1})$ parametrize a family of quantum black holes at fixed backreaction \cite{Emparan:2020znc}. It is worth distinguishing between solutions with $\kappa=+1$ and those with $\kappa=-1$. The former can have $M<0$ and are considered quantum-dressed conical defects, while the latter have $M\geq0$ and represent quantum-corrected BTZ black holes. Depending on the range of parameters, qBTZ can have pathological features, e.g., naked closed timelike curves. Dismissing such traits requires parameter restrictions, particularly $1-\tilde{a}^{2}\geq0$ and $\eta>0$ \cite{Emparan:2020znc}, though the full parameter space has not yet been fully explored.

The thermodynamics of the quantum BTZ black hole is inherited from the AdS$_{4}$ bulk black hole, with $M$, $J$, and temperature $T$ \cite{Emparan:1999fd,Emparan:2020znc}. Explicit expressions are known analytically and presented in the supplemental material. Notably, the Bekenstein-Hawking area-entropy \cite{Bekenstein:1972tm,Bekenstein:1973ur,Hawking:1974sw} of the classical AdS$_{4}$ black hole is identified to be the semiclassical generalized entropy \cite{Bekenstein:1974ax}, $S_{\text{gen}}$, accounting for both (higher-curvature) gravitational and von Neuman entropy of matter, $S_{\text{gen}}=S_{\text{grav}}+S_{\text{vN}}^{\text{mat}}$. Thus, for fixed $\ell$ and $\ell_3$, the thermodynamic first law is
\beq dM=TdS_{\text{gen}}+\Omega dJ\,,\label{eq:qbtzfirstlaw}\eeq
and is valid for any strength of backreaction, i.e., any $\nu$.

\noindent \textbf{Quantum black hole superradiance.}
As with classical black holes, superradiance of a rotating quantum black hole can plausibly occur when $\omega<m\Omega$. For quantum black holes this will only happen, however, when the generalized second law \cite{Bekenstein:1974ax} holds, $dS_{\text{gen}}\geq0$. Indeed, a test wave of frequency $\omega$ impinging a neutral rotating quantum black hole will also obey $dM/dJ=\omega/m$. The quantum black hole, e.g., qBTZ, will then change according to the semiclassical first law (\ref{eq:qbtzfirstlaw}), such that
\beq dM=\frac{\omega T}{(\omega-m\Omega)}dS_{\text{gen}}\;. \label{eq:firstlawomega}\eeq
Thus, assuming the generalized second law, if $\omega<m\Omega$ energy may be extracted from the quantum black hole, suggesting it is unstable.

The above argument is a semiclassical generalization of the one proposed by Bekenstein \cite{Bekenstein:1973mi}. Notably, the scattering wave itself need not be treated quantum mechanically. Replacement of the classical horizon area for generalized entropy is a result of the quantum matter backreacting on the geometry, independent of the test wave.

Admittedly, the reasoning leading to (\ref{eq:firstlawomega}) is only heuristic. Indeed, when backreaction is switched off, the qBTZ becomes the classical BTZ black hole, which recall does not always exhibit superradiance. A reason to expect the qBTZ black hole features superradiance opposed to its classical sibling, is that, like higher-dimensional Kerr-AdS black holes \cite{Hawking:1999dp}, there are regions which rotate faster than the speed of light. Thence, energy escaping across the horizon will be negative, allowing for superradiance to occur. We present this in the supplemental material.

\noindent \emph{Perturbing quantum BTZ.} To directly test superradiance, we perturb the rotating qBTZ with a massless probe scalar field $\Phi$ bound to the brane and analyze the associated quasinormal modes (QNMs), resonant modes of linear perturbations satisfying ingoing (outgoing) boundary conditions at the horizon (infinity) \cite{Vishveshwara:1970zz}. Our approach follows \cite{Cartwright:2024iwc}.\footnote{In~\cite{Cartwright:2024iwc} the authors consider the general QNM spectrum of the non-rotating qBTZ black hole. The general QNM spectrum of the rotating qBTZ black hole described in this work will be displayed in a companion paper.} Let the scalar field have action
\begin{equation}
    I_{\Phi}=\int \exd r \exd^2x \sqrt{-g} \partial_{a}\Phi\partial^{a}\Phi\, .
\end{equation}
Decompose the field into its Fourier modes
\begin{equation}\label{eq:mode_expansion}
\Phi(r,\bar{t},\bar{\phi})=\sum_{m= -\infty}^{\infty}\int d\omega e^{-i (\omega \bar{t} - m \bar{\phi})}\bar{\Phi}_m(\omega,r) \, .
\end{equation}
for frequency $\omega$ and azimuthal number $m$. (Here, for convenience, we work in coordinates $(\bar{t},r,\bar{\phi})$, such that the coordinate system is not identical to those given in metric (\ref{eq:qbtzrotatmetmain}).)
Solutions to the scalar equation of motion, $\Box\Phi=0$, near the asymptotic AdS$_{3}$ boundary behave as
\begin{equation}~\label{eq:scalar_field_asymp}
    \Phi\sim \Phi_{(0)}r^{-\Delta_-}+\Phi_{(+)}r^{-\Delta_+},\qquad \Delta_{\pm}=1\pm 1\,,
\end{equation}
with scalar operator dimension $\Delta_{\pm}$, and coefficients $\Phi_{(0)}$ and $\Phi_{(+)}$ are functions of $(\bar{t},\bar{\phi})$. Physically, $\Phi_{(0)}$ represents the source and $\Phi_{(+)}$ the vacuum expectation value of the scalar.  Meanwhile, near the horizon,
\begin{equation}
    \Phi \sim \Xi_1   e^{-i r_*(\omega -m \Omega )} + \Xi_2 e^{+i r_* (\omega -m \Omega )}
\label{eq:Phinearhor}\end{equation}
for tortoise coordinate $r_*$ (near the horizon)
\begin{equation}
    r_*=\frac{\log (r-r_+)}{2 \kappa_g},\quad \kappa_g=\sqrt{-\frac{1}{2}\nabla_a\chi_b\nabla^a\chi^b}
\label{eq:torcoord}\end{equation}
with surface gravity $\kappa_g$.
To obtain QNMs we use boundary conditions that disallow outgoing flux across the horizon and force the field to vanish at infinity, imposing $\Xi_2=0$ and $\Phi_{(0)}=0$, respectively.

In general, QNM frequencies are complex, $\omega=\text{Re}(\omega)+i\,\text{Im}(\omega)$, signaling the decay or growth of the field. In the supplemental material we display an example of the QNM spectrum of a scalar fluctuation with vanishing momentum ($m=0$). At $\nu=0$, the QNMs agree with the classical BTZ spectrum~\cite{Cardoso:2001hn,Birmingham:2001hc,Birmingham:2001pj}. At vanishing momentum $m$, these QNMs are purely imaginary and move along the imaginary axis as the black hole angular momentum is increased. Quantum backreaction strongly affects the QNM spectrum, and in particular causes the modes of the BTZ black hole to acquire a finite real part, hence causing a transition from purely dissipative to propagating modes. This same behavior was seen for static qBTZ~\cite{Cartwright:2024iwc}.

\noindent \textit{Superradiant modes.} Superradiance manifests itself as unstable perturbations that grow in time. According to the decomposition (\ref{eq:mode_expansion}), if the imaginary part of $\omega$ is positive, $\text{Im}(\omega)>0$, the field will grow exponentially, indicating an instability; a stable mode has $\text{Im}(\omega)<0$. Hence, the onset of the instability occurs when $\text{Im}(\omega)=0$, i.e., $\omega\in\mathbb{R}$. This can be made more precise by the following argument. Since the geometry is stationary, the field equations are invariant under $\bar{t}\rightarrow -\bar{t}$ and
$\omega\rightarrow -\omega$ hence the complex conjugate $\bar{\Phi}^*$ is also a solution to the wave equation. Indeed if one takes the tortoise coordinate
\begin{equation}
    \frac{dr_*}{dr}=\frac{H'(r_+) r_+}{2 r \kappa_g H(r)} \,
\end{equation}
then one can show that the differential equation satisfied by $\Phi$ can be put in the form
\begin{equation}
    \bar{\Phi}''(r_*)+V_{\text{eff}}(r(r_*))\bar{\Phi}(r_*)=0 \,.
\end{equation}
By Abel's identity, the Wronskian is independent of $r_{\ast}$, $W(\bar{\Phi},\bar{\Phi}^*)(r_*)=W(\bar{\Phi},\bar{\Phi}^*)$, and hence conserved along the radial direction. The Wronskians evaluated at the horizon and asymptotic boundary thus agree,  $W(\bar{\Phi},\bar{\Phi}^*)\left.\right|_{r_+}^{\infty}=0$. Using the expansions in (\ref{eq:scalar_field_asymp}) and (\ref{eq:Phinearhor}),
\begin{equation}
   \bar{\Phi}_{(0)}^*\bar{ \Phi}_{(+)}-\bar{\Phi}_{(0)} \bar{\Phi}_{(+)}^*= i (\omega -m \Omega ) |\Xi_1|^2\;,
\end{equation}
where we absorbed a constant prefactor into the definition of the source and vacuum expectation value. The Ward-Takahashi identities then imply (cf.~\cite{Ishii:2022lwc}),
\begin{equation}
\begin{split}
    \partial_t \mathscr{M}&= -2\pi\left( \Phi_{(0)}^*\partial_t \Phi_{(+)}-\Phi_{(0)} \partial_t \Phi_{(+)}^*\right) \, ,\\
    &= 4\pi\omega(\omega -m \Omega )  |\Xi_1|^2 \, ,
\end{split}
\end{equation}
where $\mathscr{M}=\int_{\Sigma}T_{ij}k^in^j$ is the energy of the spacetime on a constant time, one-dimensional spacelike hypersurface $\Sigma$, with $n^i$ the unit normal to $\Sigma$ and $k^i$ the timelike Killing vector. This implies the energy decreases, $\partial_t\mathcal{M}<0$, when the condition $\omega< m\Omega$ is fulfilled.

\begin{figure*}[t!]
    \centering
    \includegraphics[width=0.45\textwidth]{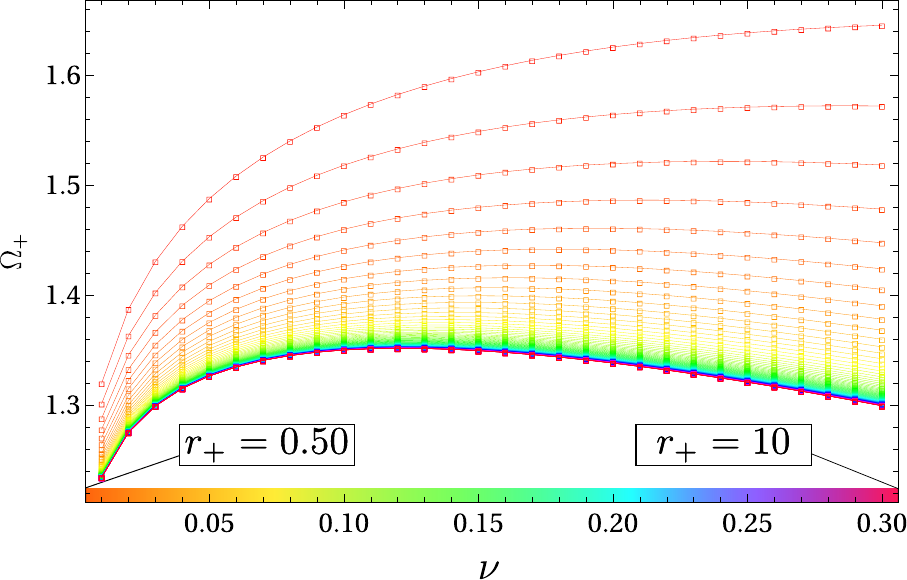}\hfill
    \includegraphics[width=0.45\textwidth]{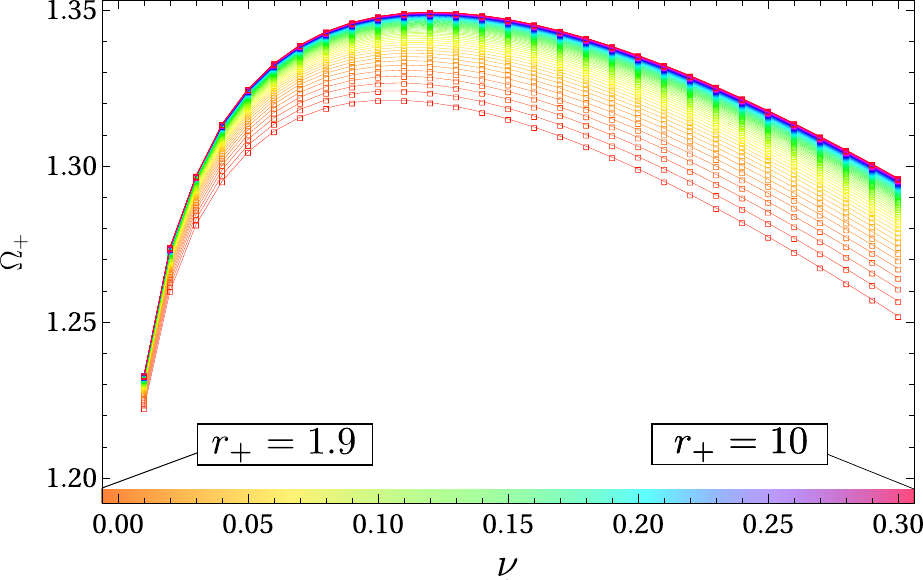}$\quad\,\,\,$
    \vspace{-3mm}
    \caption{The onset of superradiance for the rotating qBTZ black hole. \textit{Left:} Dressed cones ($\kappa=+1$). \textit{Right:} Corrected black holes ($\kappa=-1$). The color of the points represents increasing horizon radius.
    \label{fig:onset_n_1}}
\end{figure*}

To precisely obtain superradiant modes, we resort to numerics. At the onset of the instability the QNM frequency must take the form $\omega=m\Omega_+$. We solve the equation of motion for the scalar field subject to the boundary condition of no outgoing modes at the horizon with $\omega=m\Omega_{+}$. The solution near the asymptotic boundary takes the form Eq.~(\ref{eq:scalar_field_asymp}). We use a shooting method from the outer horizon and search for solutions with $\Phi_{(0)}=0$. We display the search in Fig.~\ref{fig:onset_n_1}, where we plot $\Omega_+$ as a function of $\nu$ for modes at the cusp of instability for rotating quantum dressed cones and corrected black holes.

\begin{figure}[t]
    \centering
    \includegraphics[width=0.48\textwidth]{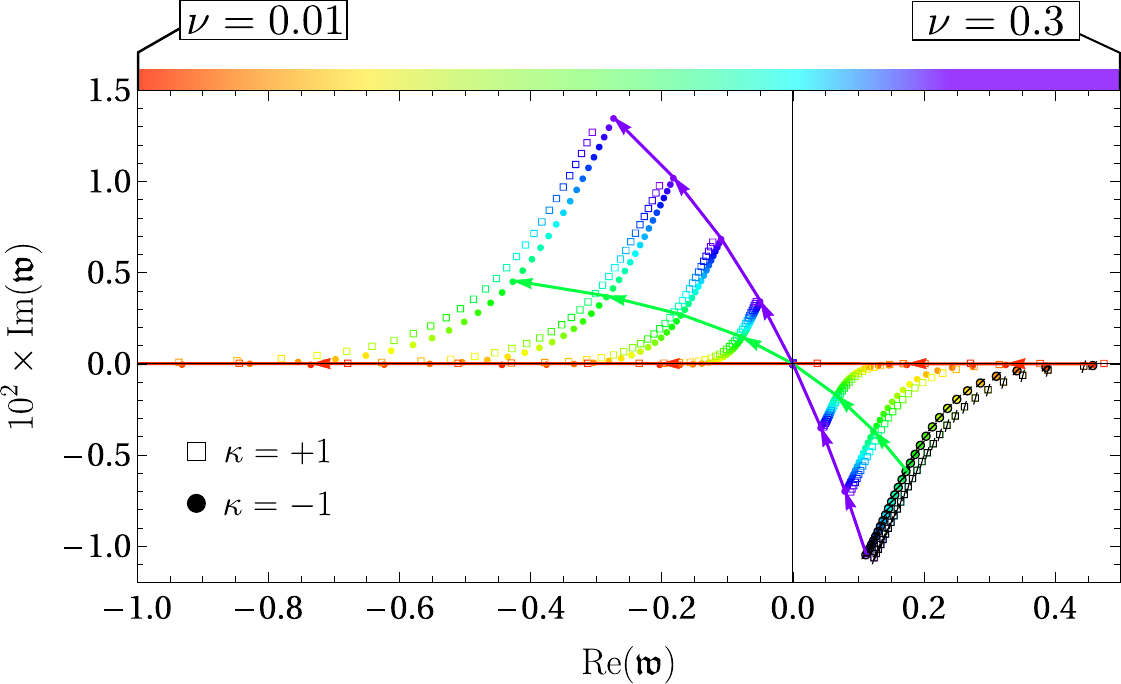}
    \vspace{-7mm}
    \caption{Superradiant QNMs ($\mathfrak{w}=\omega/(2\pi T)$) in a co-rotating frame ($\mathfrak{w}_{onset}=0$). Displayed are $m=1$ modes at $x_1=0.21$. Solid circles denotes $\kappa=-1$ while the empty boxes correspond to $\kappa=+1$, and color represents increasing strength of backreaction.
    Lines with arrows indicating increasing values of $a$ are drawn for $\nu=\{0.01,0.15,0.3\}$ to guide the eye. Note that all data for which $\text{Re}(\mathfrak{w})>0$ has $\text{Im}(\mathfrak{w})<0$.
    \label{fig:SuperRadiant_Modes}}
\end{figure}

Notably, at least for the lightest modes ($m=1$), we find no parameter combinations for which superradiant modes exist for fully non-perturbative quantum corrected BTZ black holes ($\nu>1$), and thus such black holes appear to not exhibit superradiant instabilities for the massless scalar. Such instabilities only exist for weak backreaction, $\nu<1$, see Fig. \ref{fig:SuperRadiant_Modes}. Alternatively, light superradiant modes exist for all $\nu$ for the quantum-dressed conical defects (supplemental material). Thus, while quantum-dressed cones are always dynamically unstable for the lightest modes, quantum-corrected black holes are stable against the lightest modes for large backreaction.

\noindent \emph{Quantum black hole bombs.} Repeated superradiant amplification leads to superradiant instabilities as superradiant modes become locally trapped near the black hole \cite{Press:1972zz}. For AdS black holes, this trapping naturally arises due to the timelike boundary, forming a black hole ``bomb'' for small black holes ($r_{+}/\ell_{3} \ll 1$) \cite{Cardoso:2004nk,Cardoso:2004hs,Cardoso:2006wa,Cardoso:2013pza,Chesler:2018txn}, while large ones ($r_{+}/\ell_{3} > 1$) remain stable. Likewise, the onset of superradiant instability for the qBTZ, bolstered by the existence of superradiant QNMs (Fig \ref{fig:SuperRadiant_Modes}) implies the formation of quantum black hole bombs. Notably, quantum-dressed cones and quantum-corrected black holes need not be extremely small (e.g., $r_{+}/\ell_{3}\sim .75$ in Fig. \ref{fig:onset_n_1supp},).

Superradiant instabilities are inherently dynamical; they do not necessarily coincide with thermal instability. For Kerr-AdS black holes, superradiant instability is consistent with thermal instability, as small black holes have negative heat capacity \cite{Hawking:1982dh}. It is worth then to analyze the thermal stability of the rotating qBTZ black hole. Previous investigations include evaluating only the behavior of the heat capacity \cite{Kudoh:2004ub,Frassino:2023wpc,Johnson:2023dtf,HosseiniMansoori:2024bfi,Frassino:2024bjg}.
Here we go beyond analyzing only the heat capacity, and instead study the sign of the eigenvalues of the Hessian matrix
\beq \text{Hess}_{S}=\begin{pmatrix} \frac{\partial^{2}S_{\text{gen}}}{\partial M^{2}}& \frac{\partial^{2}S_{\text{gen}}}{\partial J\partial M}\\ \frac{\partial^{2}S_{\text{gen}}}{\partial M\partial J}&\frac{\partial^{2}S_{\text{gen}}}{\partial J^{2}}\end{pmatrix}\;,\label{eq:hessian}\eeq
for entropy $S_{\text{gen}}$ (see \cite{Emparan:2020znc} for explicit expressions). Analogous to ordinary thermal systems, we expect thermal instability if the Hessian has a positive eigenvalue. We find, at the onset of instability, the two eigenvalues of Hess$_{S}$ for quantum-corrected black holes ($\kappa=-1$) are both negative, indicating a thermally stable system that exhibits dynamical instabilities. Meanwhile,  for the quantum-dressed cones ($\kappa=1$), Hess$_{S}$ has only negative eigenvalues for $M>0$, but for $M<0$ it can have at least one positive eigenvalue over a specific range of parameters. Thus, for the dressed cones there exist solutions that are both thermodynamically and dynamically unstable.

\noindent \textbf{Discussion.} We established that quantum backreaction can induce superradiance in black holes which would otherwise not exhibit the phenomenon. Here we focused on the rotating quantum BTZ black hole, finding superradiant modes giving rise to instabilities. Superradiance of course depends on the choice of positive frequency, a subtle issue for rotating black holes as they typically do not possess a globally timelike Killing vector. Here we take $\omega>0$. It is conceivable that other choices would imply no superradiance, as in Kerr-Newman-AdS$_{4}$ \cite{Winstanley:2001nx}. A more detailed mode analysis would shed more light on this subtlety. Further, a non-holographic perturbative computation of backreaction due to a conformally coupled scalar in rotating BTZ \cite{Casals:2019jfo} suggests no superradiance. Their corrected metric, however, is more complicated than the exact qBTZ \cite{Emparan:2020znc}, and evidently displays different physics. In fact, a reason why the qBTZ black hole exhibits superradiance, in contrast with the classical BTZ or the perturbatively-corrected solution, is due to the existence of a new macroscopic scale, $\ell\sim c_{3} L_{\text{P}}$ \cite{Cardoso:2004nk}. Our work has many implications and avenues worth exploring.

\noindent \emph{Beyond rotating qBTZ.} Our investigation here is non-exhaustive, being centered on classical massless probe scalars in rotating qBTZ. It would be straightforward to apply our methods to other types of quantum black holes. This includes the charged qBTZ \cite{Climent:2024nuj,Feng:2024uia}, and their charged/rotating de Sitter and Minkowski counterparts \cite{Emparan:2022ijy,Panella:2023lsi,Climent:2024wol,Panella:2024sor}. Notably, in 3D vacuum general relativity there are no de Sitter or flat black holes to begin with. Should our results hold beyond AdS$_{3}$, not only would backreaction induce black hole horizons, but all 3D quantum black holes would exhibit superradiance.

Not all higher-dimensional Kerr-Newman black holes exhibit superradiance \cite{Hawking:1999dp,Winstanley:2001nx}. It would be interesting to see if backreaction in these backgrounds would likewise facilitate superradiance. This is technically challenging because exact quantum black hole solutions (non-perturbative in backreaction) in four or higher dimensions are not known. A first step would be to study backgrounds with perturbative backreaction; indeed, our findings show that superradiance is induced for quantum-corrected black holes at weak backreaction.

Finally, it would be worth considering other types of probes, e.g., massive scalars, or conformally coupled fields. It would also be interesting to extend to quantum matter probes, holographically dual to 3+1 dimensional bulk fields~\cite{Chung:2015mna}. The quantum analog of classical superradiance is well-known for Kerr and static, charged black holes \cite{Starobinsky:1973aij,Unruh:1974bw,Gibbons:1975kk,Balakumar:2020gli}, where the black holes spontaneously emit particles with classically-superradiant modes. In principle, similar non-thermal radiation is expected for quantum test fields in a quantum black hole.

\noindent \emph{Testing cosmic censorship.} General relativity predicts black holes formed under collapse develop singularities \cite{Penrose:1964wq}. To maintain predictability, Penrose conjectured ``cosmic censorship'', which in its weak form (WCCC) states black hole singularities lie behind event horizons \cite{Penrose:1968ar,Penrose:1969pc}. Tests of WCCC, where one attempts to overspin or overcharge Kerr-Newman black holes via particle/wave scattering \cite{Wald:1974hkz,Sorce:2017dst}, have only shown WCCC is robust. In fact, scalar matter that overspins Kerr, thus violating WCCC, is repelled due to superradiance \cite{Hod:2008zza}. Similarly, the rotating qBTZ cannot be overspun \cite{Frassino:2024fin}. It would be interesting to check if the induced superradiance uncovered here is the cause.

\noindent \emph{Stability of quantum black holes.} Here we uncovered unstable superradiant modes in the qBTZ geometry. Further, at the onset of instability, the quantum-corrected black holes were found to be thermally stable, while, for specific parameter ranges the quantum-dressed cones were thermally unstable. This is consistent with the statement of Gubser-Mitra \cite{Gubser:2000mm} -- all black branes corresponding to thermodynmically unstable black holes are also dynamically unstable -- and suggests the mantra extends semiclassically. For classical stationary and axisymmetric black holes, dynamical stability is equivalent to the following criterion \cite{Hollands:2012sf}: non-negativity of the canonical energy $\mathcal{E}$, a bilinear form over linearized perturbations that fix appropriate conserved quantities. Conversely, for AdS black holes, $\mathcal{E}<0$ indicates a superradiant instability \cite{Green:2015kur}. It would be interesting to develop a similar stability criterion for quantum black holes. Naively, the classical energy and stability criterion are to be replaced with its semiclassical generalization,
\beq \mathcal{E}_{\text{sc}}=\delta^{2}M-\Omega\delta^{2}J-T\delta^{2}S_{\text{gen}}\;, \quad \mathcal{E}_{\text{sc}}\geq0\;,
\label{eq:stabcriterion}\eeq
where, as in the thermodynamic first law, the classical event horizon area has been supplanted with generalized entropy, and $\delta^{2}$ indicates second-order variations.
In fact, since the AdS C-metric belongs to the class of metrics the classical stability criterion \cite{Green:2015kur} applies to, the criterion (\ref{eq:stabcriterion}) automatically follows for the quantum BTZ black hole due to the identification of the bulk area-entropy and brane generalized entropy (as in the first law).

Weak cosmic censorship and black hole stability, moreover, are intimately linked. Indeed,
$\mathcal{E}>0$ implies the satisfaction of a local Penrose inequality (PI) \cite{Hollands:2012sf}, an inequality whose violation yields a counterexample to WCCC \cite{Penrose:1973um}. The local PI is thus a necessary and sufficient condition for dynamical stability. Quantum matter, however, is known to violate the classical Penrose inequality \cite{Bousso:2019bkg,Bousso:2019var}. This motivates the need for (weak) quantum cosmic censorship,\footnote{Such a notion was formulated in \cite{Engelhardt:2024hpe}.} for which a quantum Penrose inequality would be an input. To date, all quantum BTZ black holes obey such a proposed inequality \cite{Frassino:2024bjg} for all orders of backreaction. It would be interesting see how these robust quantum inequalities relate to the proposed stability criterion (\ref{eq:stabcriterion}).

\noindent \emph{Acknowledgements.}
We are grateful to Vitor Cardoso, Roberto Emparan and Robie Hennigar for useful correspondence. CC and UG are supported by the Netherlands Organisation for Scientific Research (NWO) under the VICI grant VI.C.202.104. Work of UG is also supported by the ENW-XL program “Probing the QCD phase diagram” of NWO. JFP is supported by the `Atracci\'on de Talento' program grant 2020-T1/TIC-20495, by the Spanish Research Agency through the grants CEX2020-001007-S and PID2021-123017NB-I00, funded by MCIN/AEI/10.13039/501100011033, and by ERDF A way of making Europe. AS is supported by STFC grant ST/X000753/1.

\bibliographystyle{apsrev4-2}
\bibliography{superradbib}

\newpage

\onecolumngrid

\section{Supplemental material}

\begin{center}
\textbf{Braneworld framework}
\end{center}

\noindent Exact descriptions of three-dimensional braneworld black holes \cite{Emparan:1999wa,Emparan:1999fd} follow from embedding an end-of-the-world (ETW) brane in a particular AdS$_{4}$ C-metric. In Boyer-Lindquist coordinates the line element of the neutral, rotating AdS$_{4}$ C-metric is
\begin{equation}
\begin{split}
 ds^2 = \frac{\ell^2}{(\ell + xr)^2} \biggr[ & -\frac{H(r)}{\Sigma(x,r)} \left(dt+ax^2 d\phi \right)^2 + \frac{\Sigma(x,r)}{H(r)} dr^2 + r^2 \left(\frac{\Sigma(x,r)}{G(x)}dx^2 + \frac{G(x)}{\Sigma(x,r)}\left( d\phi- \frac{a}{r^2} dt \right)^2  \right) \biggr] \ ,
 \end{split}
\label{eq:rotatingCmetqbtz}\end{equation}
with metric functions
\beq
\begin{split}
    &H(r)= \frac{r^2}{\ell_3^2}+ \kappa -\frac{\mu \ell}{r}+ \frac{a^2}{r^2}  \ , \quad G(x)= 1-\kappa x^2-\mu x^3+ \frac{a^2}{\ell_3^2}x^4 \ , \\
   &\Sigma(x,r)= 1 + \frac{a^2 x^2}{r^2} \ .
\end{split}
\label{eq:metfuncsrotqbtz}\eeq
This geometry may be interpreted as describing a single or pair of uniformly accelerating black holes due to a cosmic string or strut (see, e.g., \cite{Griffiths:2006tk}) with (inverse) acceleration $\ell$. The metric is a solution to pure Einstein-AdS$_{4}$ gravity,
\beq I=\frac{1}{16\pi G_{4}}\int d^{4}x\sqrt{-\hat{g}}\left[\hat{R}+\frac{6}{\ell_{4}^{2}}\right]\;, \eeq
where $\ell_{4}$ denotes the AdS$_{4}$ radius, related to parameters $\ell_{3}$ and $\ell$ via
\beq \frac{1}{\ell_{4}^{2}}=\frac{1}{\ell_{3}^{2}}+\frac{1}{\ell^{2}}\;.\eeq
with $0<\ell<\infty$. Further,  $\kappa=\pm1,0$ corresponds to different slicings of the boundary ($\kappa=-1$ results in BTZ black holes), $\mu$ is a mass parameter, and the non-negative parameter $a$ introduces rotational effects.

In the bulk geometry (\ref{eq:rotatingCmetqbtz}), roots $r_{i}$ of $H(r)$ occur when the Killing vector
\beq \chi^{b}=\partial^{b}_{t}+\frac{a}{r_{i}^{2}}\partial^{b}_{\phi}\;,\label{eq:genzetabtz}\eeq
for $r_{i}$ finite, has vanishing modulus, $\chi^{2}=g_{ab}\chi^{a}\chi^{b}=-\frac{\ell^{2}}{(\ell+x r)^{2}}H(r_{i})\Sigma(x,r_{i})=0$ at $r=r_{i}$. The largest positive, real root of $H(r)$, denoted $r_{+}$, corresponds to the radius of the outer event horizon of the black hole, while the inner black hole horizon is denoted by $r_{-}$ (where $r_{-}< r_{+}$). In the limit of vanishing rotation $a\to0$, then $r_{-}\to0$.

The real zeroes $x_{i}$ of $G(x)$ give rise to conical singularities on the black hole event horizon. These conical defects result in, for a example, a cosmic string suspending a single black hole away from the center of AdS$_{4}$, resulting in its acceleration. One of these conical singularities can be removed by imposing regularity to ensure smoothness of the geometry along the axis of rotational symmetry,
\beq \phi\sim \phi+\Delta\phi\;,\quad \Delta\phi=\frac{4\pi}{|G'(x_{i})|}\;.\label{eq:bulkregcondphi}\eeq
In these constructions the smallest positive root, denoted $x=x_{1}$, is chosen, leaving conical singularities at the remaining zeroes $x_{i}\neq x_{1}$. It is standard to then restrict to the region $0\leq x\leq x_{1}$, where there are no remaining conical singularities (the other conical singularities live in the range $x<0$). The range of $x_{1}$ depends on $\mu$ and $a$.

A key geometric feature of the C-metric (\ref{eq:rotatingCmetqbtz}) is that the hypersurface $x=0$ is umbilic, i.e., extrinsic curvature $K_{ij}$ is proportional to the induced metric $h_{ij}$ at $x=0$; $K_{ij}=-\ell^{-1}h_{ij}$. Thus, a codimension-1 brane $\mathcal{B}$ at $x=0$ obeys the Israel junction conditions. For a purely tensional brane, with action
\beq I_{\text{brane}}=-\tau\int_{\mathcal{B}}d^{3}x\sqrt{-h}\;,\label{eq:braneactapp}\eeq
the Israel junction conditions set the tension $\tau=(2\pi G_{4}\ell)^{-1}$.
Treating the $x=0$ hypersurface as an ETW brane, the $x<0$ region is cutoff from the remainder of bulk AdS$_{4}$. The space can be surgically completed by gluing a second copy of the spacetime along $x=0$, resulting in a $\mathbb{Z}_{2}$-symmetric double-sided braneworld without a cosmic string \cite{Randall:1999vf,Karch:2000ct}.

According to braneworld holography \cite{deHaro:2000wj}, the induced action on the brane characterizes a specific semiclassical theory of gravity with an infinite tower of higher-derivative terms, coupled to a  three-dimensional conformal field theory with large central charge $c_{3}$ and an ultraviolet cutoff $\ell$ (see \cite{Emparan:2023dxm,Panella:2024sor} for a pedagogical summary). The precise form of the brane-induced action can be found in, e.g., \cite{Emparan:2020znc,Panella:2024sor}. Pertinently, the effective couplings on the brane are induced from the four-dimensional parent theory
couplings $\{G_{4},\ell_{4},\tau\}$
\beq
G_{3}=\frac{1}{2\ell_{4}}G_{4} \,,\quad \frac{1}{L_{3}^2}=\frac{2}{\ell_{4}^2}\left(1-2\pi G_{4}\ell_{4}\tau\right)\,.
\eeq
Here $L_3$ is the effective AdS$_{3}$ radius appearing in the induced brane cosmological constant. For small backreaction ($\nu\ll1$), $L_{3}$ approximately equals the curvature radius $\ell_{3}$.

\begin{center}
\textbf{Quantum BTZ solution}
\end{center}

We summarize relevant details of the neutral rotating quantum BTZ black hole (see \cite{Emparan:2020znc,Panella:2024sor} for more details).

\vspace{2mm}

\noindent \emph{Geometry.} The line element (\ref{eq:rotatingCmetqbtz}) at $x=0$ gives the `naive metric'
\beq ds^{2}|_{x=0}=-H(r)dt^{2}+H^{-1}(r)dr^{2}+r^{2}\left(d\phi-\frac{a}{r^{2}}dt\right)^{2}\;,\label{eq:naivemetapp}\eeq
with $H(r)$ as in (\ref{eq:metfuncsrotqbtz})
The metric is naive because the bulk regularity condition (\ref{eq:bulkregcondphi}) removing the conical singularity has not yet been invoked. Removal of the singularity requires one simultaneously perform the coordinate transformation $(t,\phi)\to(\bar{t},\bar{\phi})$ \cite{Emparan:2020znc} (see section 4.2.2 of \cite{Panella:2024sor} for more details)
\beq t=\eta(\bar{t}-\tilde{a}\ell_{3}\bar{\phi})\;,\quad \phi=\eta\left(\bar{\phi}-\frac{\tilde{a}}{\ell_{3}}\bar{t}\right)\;,\label{eq:tbarphibar}\eeq
where
\beq \eta\equiv\frac{\Delta\phi}{2\pi}=\frac{2x_{1}}{3-\kappa x_{1}^{2}-\tilde{a}^{2}}\;,\eeq
with $\tilde{a}\equiv ax_{1}^{2}/\ell_{3}$. The Killing vectors $\partial_{t}$ and $\partial_{\phi}$ transform as
\beq \partial_{t}=\frac{1}{\eta(1-\tilde{a}^{2})}\left(\partial_{\bar{t}}+\frac{\tilde{a}}{\ell_{3}}\partial_{\bar{\phi}}\right)\;,\quad \partial_{\phi}=\frac{1}{\eta(1-\tilde{a}^{2})}\left(\partial_{\bar{\phi}}+\tilde{a}\ell_{3}\partial_{\bar{t}}\right)\;.\eeq
In coordinates (\ref{eq:tbarphibar}) points are identified along orbits of $\partial/\partial\bar{\phi}$ with $(\bar{t},\bar{\phi})\sim(\bar{t},\bar{\phi}+2\pi)$.

Though bulk regularity has been imposed, the metric (\ref{eq:naivemetapp}) in terms of coordinates (\ref{eq:tbarphibar}) is not quite in a canonical asymptotic form for a rotating AdS black hole. A canonically normalized radial coordinate $\bar{r}$ related to $r$ is often further introduced
\beq r^{2}\equiv \frac{\bar{r}^{2}-r_{s}^{2}}{(1-\tilde{a}^{2})\eta^{2}}\;,\qquad r_{s}=2\tilde{a}\ell_{3}\frac{\sqrt{2-\kappa x_{1}^{2}}}{3-\kappa x_{1}^{2}-\tilde{a}^{2}}=\frac{\tilde{a}\ell_{3}\eta}{x_{1}}\sqrt{2-\kappa x_{1}^{2}}\;,\eeq
where the $\bar{r}=r_{s}$ denotes the location of the ring singularity (at $r=0$ in the original coordinates).
In canonically normalized coordinates $(\bar{t},\bar{r},\bar{\phi})$, the rotating quantum BTZ black hole has line element \cite{Emparan:2020znc}
\beq
\hspace{-6mm}
\begin{split}
ds^2 = &-\left(\frac{\bar{r}^2}{\ell_3^2}-8\mathcal{G}_3 M -\frac{\ell \mu \eta^2}{r} \right) d\bar{t}^2 + \left(\frac{\bar{r}^2}{\ell_3^2}-8\mathcal{G}_3 M + \frac{(4 \mathcal{G}_3 J)^2}{\bar{r}^2}- \ell \mu (1-\tilde{a}^2)^2 \eta^4 \frac{r}{\bar{r}^2} \right)^{\hspace{-2mm}-1} \hspace{-1mm}d\bar{r}^2  \\
& + \left(\bar{r}^2 + \frac{\mu \ell\tilde{a}^2 \ell_3^2 \eta^2}{r} \right) d\bar{\phi}^2 - 8 \mathcal{G}_3 J \left( 1+ \frac{\ell}{x_1 r} \right) d\bar{\phi} d\bar{t}\;,
\end{split}
\label{eq:qbtzrotatmet}\eeq
where both $r$ and $\bar{r}$ (treating $r=r(\bar{r})$) have been kept when for convenience. Here $M$ and $J$ are interpreted as the mass and angular momentum of the braneworld black hole,
\beq
\begin{split}
8 \mathcal{G}_3 M &= -\kappa\eta^2 \left(1+ \tilde{a}^2-\frac{4 \tilde{a}^2}{\kappa x_1^2} \right) = 4 \frac{- \kappa x_1^2 + \tilde{a}^2(4-\kappa x_1^2)}{(3-\kappa x_1^2 - \tilde{a}^2)^2}
\end{split}
\label{eq:massrotqbtz}\eeq
\beq
\begin{split}
4 \mathcal{G}_3 J &= \ell_3\eta^2 \tilde{a}\mu x_1=\frac{4 \ell_{3}\tilde{a}(1-\kappa x_1^2 + \tilde{a}^2)}{(3 - \kappa x_1 ^2 -\tilde{a}^2)^2}\;,
\end{split}
\label{eq:rotJqbtz}\eeq
and obey
\beq 8\mathcal{G}_{3}\left(M\pm \frac{J}{\ell_{3}}\right)=\frac{4(1-\tilde{a}^{2})(-\kappa x_{1}^{2}+2\tilde{a}^{2})}{(3-\kappa x_{1}^{2}-\tilde{a}^{2})^{2}}\;.\eeq
Here it is used that $\mu=(1-\kappa x_{1}^{2}+\tilde{a}^{2})/x_{1}^{3}$. Notice one recovers the classical rotating BTZ geometry when $\ell=0$ in the metric (\ref{eq:qbtzrotatmet}).

In coordinates $(\bar{t},\bar{r},\bar{\phi})$, the inner and outer horizons of the quantum black hole, $\bar{r}_{-}$ and $\bar{r}_{+}$, are generated by orbits of the canonically normalized generator (\ref{eq:genzetabtz}),
\beq \bar{\chi}^{b}_{\pm}\equiv \frac{\eta(1-\tilde{a}^{2})}{1+\frac{a^{2}x_{1}^{2}}{r_{\pm}}}\zeta^{b}=\partial_{\bar{t}}+\Omega_{\pm}\partial_{\bar{\phi}}\;.\eeq
Here $\Omega_{\pm}$ refers to the angular velocity of the horizons $r_{\pm}$ relative to a \emph{non-rotating} frame at spatial infinity
\beq \Omega_{\pm}\equiv \frac{a}{r_{\pm}^{2}+a^{2}x_{1}^{2}}\left(1+\frac{r_{\pm}^{2}x_{1}^{2}}{\ell_{3}^{2}}\right)\;.\label{eq:Omqbtznonrotinf}\eeq
The angular velocity $\Omega'_{\pm}$ relative to a \emph{rotating} frame at spatial infinity is
\beq \Omega'_{\pm}\equiv\frac{a}{r_{\pm}^{2}+a^{2}x_{1}^{2}}\left(1-\frac{a^{2}x_{1}^{4}}{\ell_{3}^{2}}\right)\;,\eeq
obeying $\Omega_{\pm}-\Omega'_{\pm}=ax_{1}^{2}/\ell_{3}^{2}$. Further, relative to $\bar{\chi}^{b}$, the surface gravities of the inner and outer horizons, defined via $\bar{\chi}^{b}\nabla_{b}\bar{\chi}^{c}\equiv\kappa_{\pm} \bar{\chi}^{c}$, are
\beq \kappa_{\pm}=\frac{\eta(1-\tilde{a}^{2})}{(r_{\pm}^{2}+a^{2}x_{1}^{2})}\frac{r_{\pm}^{2}}{2}|H'(r_{\pm})|\;.\eeq
In the extremal limit, $r_{+}=r_{-}$, the surface gravities $\kappa_{\pm}$ vanish.

The ergoregion corresponds to the stationary limit surface characterized by the locus of points $g_{ab}\partial^{a}_{\bar{t}}\partial^{b}_{\bar{t}}=0$, i.e., at $\bar{r}=\bar{r}_{\text{ergo}}$ where the $\bar{t}\bar{t}$-component of the metric (\ref{eq:qbtzrotatmet}) vanishes
\beq \frac{\bar{r}^{2}}{\ell_{3}^{2}}-8\mathcal{G}_{3}M-\frac{\ell\mu\eta^{2}}{r}=0\;.\eeq
In the limit of vanishing backreaction $\ell=0$, the ergoregion lies at $\bar{r}_{\text{ergo}}=\ell_{3}\sqrt{8G_{3}M}$, as for classical BTZ.

\vspace{2mm}

\noindent \emph{Faster than light surfaces.}
Consider the corotating Killing vector $\chi$ generating the outer black hole horizon. Its norm near infinity is
\beq \chi^{2}|_{\bar{r}\to\infty}=\frac{\bar{r}^{2}}{\ell_{3}^{2}}(\Omega_{+}^{2}\ell_{3}^{2}-1)+8\mathcal{G}_{3}(M-\Omega_{+}J)+\mathcal{O}(\bar{r}^{-1})\;.\eeq
Thus, like higher-dimensional Kerr-AdS black holes, $\chi$ will be timelike everywhere outside the horizon provided $|\Omega_{+}|\ell_{3}<1$.
Alternatively, for $|\Omega_{+}|\ell_{3}>1$, then $\chi$ becomes spacelike near infinity, i.e., there is a region which rotates faster than the speed of light.  By contrast, $\chi$ is timelike outside the horizon for the classical BTZ black hole: $\chi^{2}_{\text{BTZ}}=-\frac{\left(\bar{r}^2-\bar{r}_+^2\right) \left(\bar{r}_+^2-\bar{r}_-^2\right)}{\ell_3^2 \bar{r}_+^2} <0$ when $\bar{r}>\bar{r}_{+}$.

Let us see if $|\Omega_{+}|\ell_{3}>1$ is possible for qBTZ. To hold, $\Omega_+$ necessarily has a maximum for some $a$. Intuitively, this follows from the bounding behavior of $\Omega_+(a)^2$, for which $\Omega_+(0)=0$ and $\Omega_+(a=\ell/x_1^2)=1/\ell_3^2$. We find such a value of $a$ exists, given a particular $r_{+}$ and $x_{1}$, as we now show.
We look for points which extremize the horizon radius $r_{+}$ as a function of the rotation parameter $a$, i.e., when $\frac{dr_{+}}{da}\equiv r'_{+}(a)=0$. To find this condition, treat $r_{+}=r_{+}(a)$ and differentiate $H(r_{+})=0$ with respect to $a$, leading to
\beq 0=\frac{dH(r_{+}(a))}{da}=r'_{+}\left(2\kappa r_{+}-\mu\ell +\frac{4r_{+}^{3}}{\ell_{3}^{2}}\right)+2a\left(1-\frac{\ell x_{1}r_{+}}{\ell_{3}^{2}}\right)\;.\eeq
The final term is eliminated either when $a=0$ or when $r_{+}=\ell_{3}^{2}/\ell x_{1}$. Then, trivially $r'_{+}=0$. Inserting this value of $r_{+}$ into the bulk AdS$_{4}$ $H(r)$ blackening factor yields
\begin{equation}
    \frac{\left(\ell^2+\ell_3^2\right) \left(\ell^2 \left(\kappa  x_1^2-1\right)+\ell_3^2\right)}{\ell^4 x_1^4}=0\;.
\end{equation}
Solving for $x_{1}$ gives $x_1=\pm \frac{\sqrt{\ell^2-\ell_3^2}}{\ell \sqrt{\kappa}}$. We now look for extrema of $\Omega_{+}$ (\ref{eq:Omqbtznonrotinf}) as a function of $a$, using the above values of $r_{+}$ and $x_{1}$. We find
\begin{equation}
  0= \frac{\exd\Omega_{+}}{\exd a} =-\frac{\kappa  \left(\ell ^4-\ell _3^4\right) \left(a^2 \left(\ell ^2-\ell _3^2\right){}^2-\kappa ^2 \ell ^2 \ell _3^4\right)}{\left(a^2 \left(\ell ^2-\ell _3^2\right){}^2+\kappa ^2 \ell ^2 \ell _3^4\right){}^2}\;.
\end{equation}
This is satisfied either when $\ell=\ell_3$ (in which case $x_1=0$), or $a=\pm \frac{\kappa  \ell  \ell _3^2}{\sqrt{\ell ^4-2 \ell _3^2 \ell ^2+\ell _3^4}}$. Substituting in the value of $r_{+},x_{1}$ and $a$ into the angular velocity (\ref{eq:Omnonrotinf}) gives
\begin{equation}
  (\ell_3^2\Omega_+^2 -1)= \frac{\left(1+\nu^{2}\right)^2}{4\nu^2}-1\;,
\label{eq:condonOmega}\end{equation}
 a quartic polynomial in $\nu=\ell/\ell_3$ with double roots at $\nu=\pm1$. Importantly, for $\nu<1$ and $\nu>1$ we have that $(\ell_{3}^{2}\Omega_{+}^{2}-1)>0$, i.e., the existence of regions rotating faster than the speed of light. These two windows of superradiance occur for different situations: (i) ($\kappa=1$, $\nu>1$) and (ii) ($\kappa=-1$, $\nu<1$); otherwise the solution for $x_1$ becomes imaginary. Summarily, we have argued that the qBTZ can exhibit superradiance.

It is important to note that these are \textit{not} the only regimes where the Killing vector fails to be well defined near the asymptotic boundary, however, they are regimes where it is possible to explicitly show this analytically. This set of points are precisely where the angular velocity of the horizon is an extremum as a function of $a$.

\vspace{2mm}

\noindent \emph{Family of quantum black holes.} The neutral, non-rotating solution ($J=0$) comprises a family of quantum black holes with three branches labeled as 1a, 1b, and 2. Branch 1a has $\kappa=+1$ and describes black holes with non-positive mass. Branches 1b and 2 have $\kappa=-1$ and describe non-negative mass black holes. Branches 1a and 1b smoothly connect to each other (a feature that notably does not appear for the classical BTZ geometry; there is a gap in the mass spectrum of classical BTZ.) meanwhile, branches 1b and 2 meet at an upper bound on the mass.

With rotation ($J\neq0$) there is an analogous set of three branches. Now branches 1b and 2 meet at a maximum value of $M$ for fixed $J$:
\beq x_{1}^{2}+\tilde{a}^{2}=3\;,\quad M=\frac{1}{8\mathcal{G}_{3}}\left(\frac{12}{x_{1}^{4}}-1\right)\;,\quad J=\frac{\ell_{3}}{\mathcal{G}_{3}}\frac{\sqrt{3-x_{1}^{2}}}{x_{1}^{4}}\;.\label{eq:MJextbranch1}\eeq
At $x_{1}=\sqrt{2}$, one attains an extremal bound, where $M=J/\ell_{3}=1/4\mathcal{G}_{3}$. Further, there is a second extremal bound among the branch 2 black holes, found by minimizing the mass $M$ for fixed $J$:
\beq \tilde{a}=1\;,\quad M=\frac{J}{\ell_{3}}=\frac{1}{\mathcal{G}_{3}(2+x_{1}^{2})}\;,\label{eq:classextlim}\eeq
The two extremal bounds coincide at $x_{1}=\sqrt{2}$.  Recall that the classical rotating BTZ black hole obeys the extremality bound $M\geq J/\ell_{3}$. For any value of $J$, this classical extremality bound will be violated in the quantum case, $M\leq J/\ell_{3}$, when $-\kappa x_{1}^{2}<2\tilde{a}^{2}$. Such `super-extremal' black holes live among the branch 1 solutions.

\vspace{2mm}

\noindent \emph{Horizon thermodynamics.} The standard thermodynamic quantities $\{M,T,S_{\text{gen}},J,\Omega\}$ were reported in \cite{Emparan:1999fd,Emparan:2020znc}, while extended thermodynamic quantities (where the induced brane cosmological constant is treated as a dynamical thermodynamic pressure \cite{Frassino:2022zaz}) were reported in the supplemental material of \cite{Frassino:2024bjg}.
Additional restrictions on the parameter space can be imposed upon when exploring the standard thermodynamics \cite{Emparan:2020znc} (we reiterate the summary from \cite{Frassino:2024bjg}). For non-extremal solutions one takes
\beq 0\leq \alpha^{2}\leq \frac{1+\nu z}{z(z-\nu)}\;.\eeq
The lower bound is chosen to avoid naked closed timelike curves, while
the upper bound follows by requiring the outer black hole horizon radius $r_{+}$ be real and positive. Solutions with $\kappa=+1$ (solutions which negative mass conical defects which have been dressed by a horizon due to backreaction) have  $\alpha^{2}<0$. For solutions with $\kappa=-1$ and $\nu<z<\nu^{-1/3}$, the upper bound implies $1+\alpha^{2}(1-z^{2})>0$ and $\mu>0$. Notice, this applies only to a restricted range of parameters $(a,x_1,\ell,\ell_3)$. Superradiance occurs outside of this range. Even with this bound the black hole temperature can still become negative without further restriction on the range of parameters \cite{Frassino:2024bjg}.
\begin{figure}[t]
    \centering
    \includegraphics[width=0.6\textwidth]{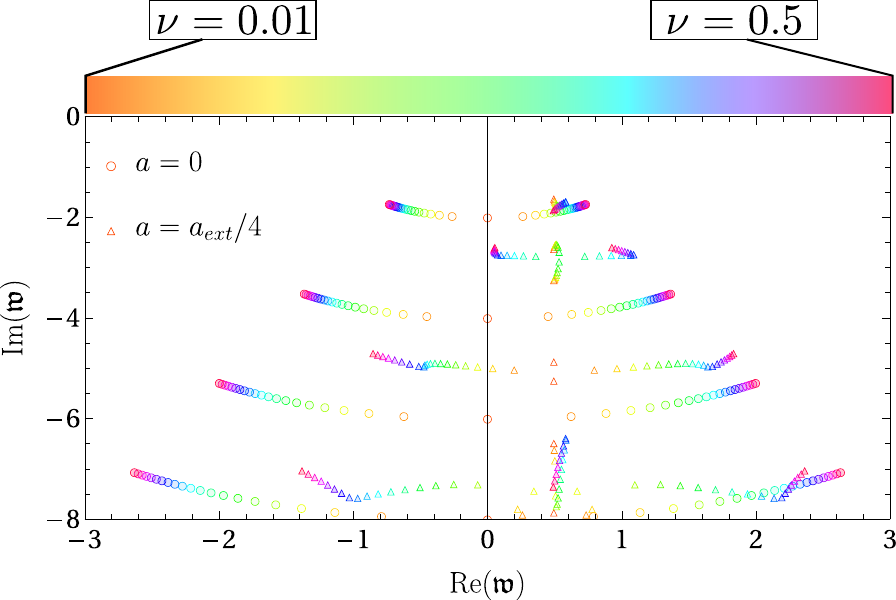}
    \caption{\textit{QNM spectrum of rotating qBTZ solution}. The colors represent the backreaction parameter $\nu$ in the range $[0,0.5]$, while the plot marker size corresponds to increasing size of rotational parameter $\tilde{a}$ in the range $[0,\tilde{a}_{ex}/4]$ with $\tilde{a}_{ex}$ being the classical extremality bound (see supplemental material). Here we take $\kappa=-1$ and fix $x_1=1$ with $\mathfrak{w}=\omega/2\pi T$.
    \label{fig:QNM}}
\end{figure}

The extremal limit of classical BTZ (\ref{eq:classextlim}) occurs when
\beq \alpha^{2}= \alpha^{2}_{\text{ext}}\equiv \frac{z^{2}(1+\nu z)}{1-2\nu z^{3}+z^{4}}\;,\eeq
where the temperature $T=0$. Since $T$ or $J$ are generally non-monotonic with respect to $\alpha$, rotating quantum black holes can exist outside the classical extremality bound.  This leads to two distinct classes: (i) sub-extremal qBTZ with $0\leq\alpha\leq \alpha_{\text{ext}}$ and (ii) super-extremal qBTZ with $\alpha> \alpha_{\text{ext}}$. Super-extremal black holes are allowed due to the combined, non-linear effects of rotation and quantum backreaction.

\vspace{2mm}

\noindent \emph{Evaluating the Hessian.} In the main text we compute the eigenvalues of the Hessian
\beq \text{Hess}_{S_{\text{gen}}}=\begin{pmatrix} \frac{\partial^{2}S_{\text{gen}}}{\partial M^{2}}& \frac{\partial^{2}S_{\text{gen}}}{\partial J\partial M}\\ \frac{\partial^{2}S_{\text{gen}}}{\partial M\partial J}&\frac{\partial^{2}S_{\text{gen}}}{\partial J^{2}}\end{pmatrix}\;,\quad (\text{Hess}_{S_{\text{gen}}})_{i,j}=\frac{\partial^{2}S_{\text{gen}}}{\partial x^{i}\partial x^{j}}\;,\label{eq:hessiansupp}\eeq
 for $x^{i}=\{M,J\}$, and $S_{\text{gen}}(M,J)$. Analogous to ordinary thermal systems, we expect thermal instability if the Hessian (\ref{eq:hessiansupp}) has a positive eigenvalue. Indeed, the Schwarzschild black hole has negative heat capacity and hence $\partial^{2}S/\partial M^{2}>0$. (If the Hessian is cast in the $M$ representation, where internal energy $M$ and entropy $S_{\text{gen}}$ swap roles, negative eigenvalues indicate thermal instability.) In practice it is non-trivial to compute elements of the Hessian (\ref{eq:hessiansupp}) for the quantum BTZ black hole because $S_{\text{gen}}$ cannot be cast purely in terms of extensive variables $M$ and $J$. Rather, we have explicit expressions for $S,M,J$ in terms of parameters $(x_{1},a)$, see Eqs. (\ref{eq:massrotqbtz}), (\ref{eq:rotJqbtz}), and \cite{Emparan:2020znc}
\beq S_{\text{gen}}=\frac{\pi}{2\ell\mathcal{G}_{3}}\frac{2\pi\ell x_{1}^{2}(r_{+}^{2}+a^{2}x_{1}^{2})}{(\ell+r_{+}x_{1})\left(3-\frac{a^{2}x_{1}^{4}}{\ell_{3}^{2}}-\kappa x_{1}^{2}\right)}\;.\label{eq:entropyqbtz}\eeq
The horizon radius $r_{+}$ itself is a function of $(x_{1},a)$, upon solving $H(r_{+})=0$. Thus, to evaluate each of the elements of the Hessian we implement the following chain rule
 \beq \frac{\partial^{2}S}{\partial x^{i}\partial x^{j}}=\frac{\partial}{\partial x^{i}}\left(\frac{\partial y^{b}}{\partial x^{j}}\frac{\partial S}{\partial y^{b}}\right)=\frac{\partial^{2}y^{b}}{\partial x^{i}\partial x^{j}}\frac{\partial S}{\partial y^{b}}+\frac{\partial y^{b}}{\partial x^{i}}\frac{\partial y^{c}}{\partial x^{j}}\frac{\partial^{2}S}{\partial y^{b}\partial y^{c}}\;,\eeq
 where we drop the subscript on $S_{\text{gen}}$ for notational simplicity, and introduce $y^{b}=\{x_{1},a\}$ (we work in thermal ensembles where $\ell$ and $\ell_{3}$ are kept fixed). Since we do not have explicit expressions for $a=a(M,J)$ and $x_{1}=x_{1}(M,J)$, we must compute the derivatives of the parameters implicitly. Note that in evaluating the thermodynamics of the quantum BTZ black holes one often uses the $(\nu,\alpha,z)$ parametrization described above. To directly compare to parameter ranges used in computing the quasinormal modes, we found this parametrization insufficient.

\begin{figure}[t]
    \centering
    \includegraphics[width=0.6\textwidth]{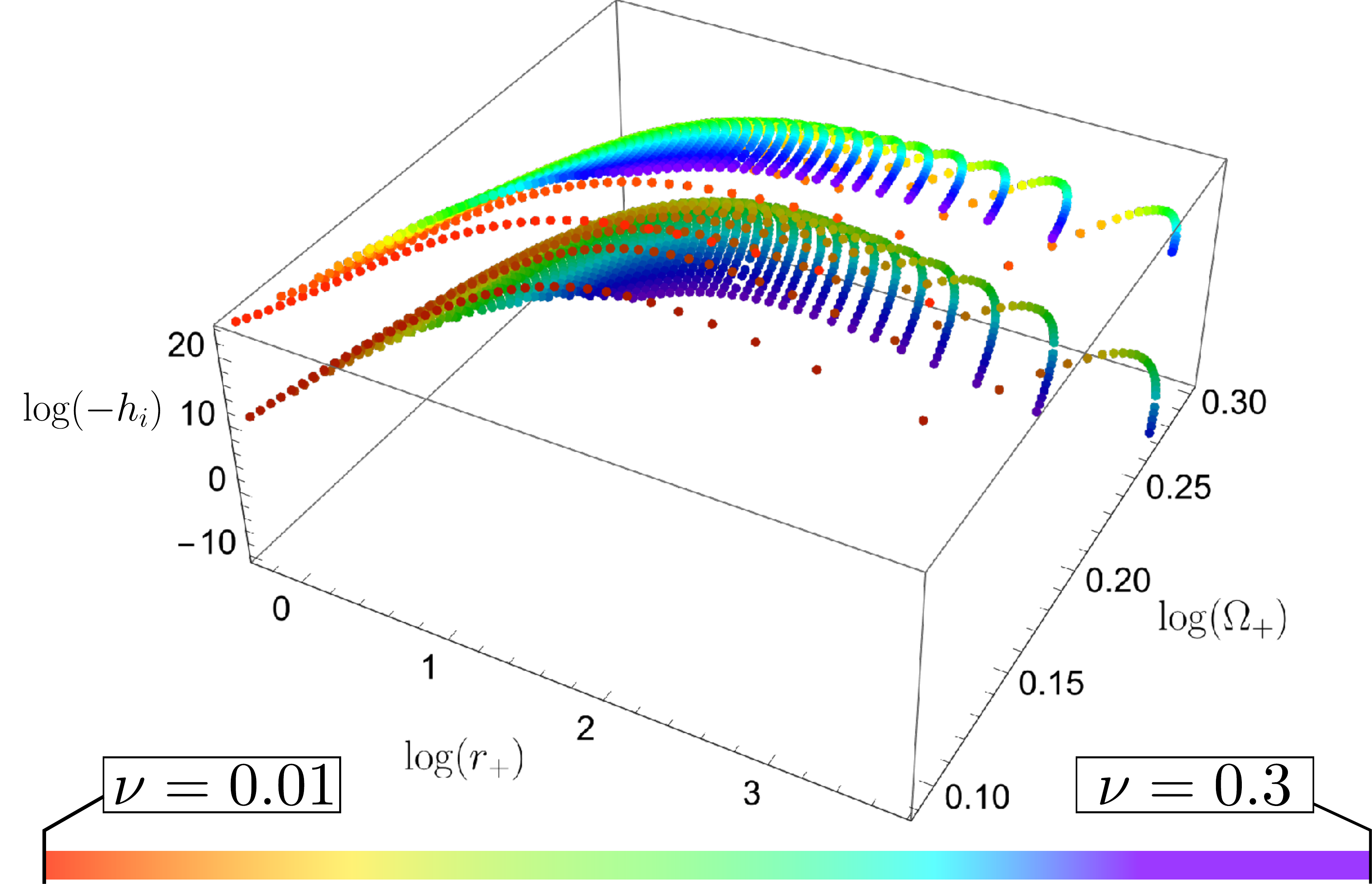}
    \caption{\textit{Eigenvalues of the Hessian}. The eigenvalues, $h_i$ for $i=1,2$ of the Hessian matrix are displayed for $\kappa=-1$. Since all eigenvalues are negative we display $\log(-h_i)$. The colors represent the backreaction parameter $\nu$ in the range $[0,0.3]$, lighter hue is $h_1$, darker hue is $h_2$. The eigenvalues displayed were computed from all the data collected at the onset of the instability, visually displayed in the right image of Fig.~\ref{fig:onset_n_1supp}.
    \label{fig:Eigen}}
\end{figure}

 Let us now briefly outline the steps needed to compute each of the elements of the Hessian matrix (we share our ancillary Mathematica files for additional aid). (1) We first implicitly determine first and second order derivatives of $r_{+}$ with respect to $a$ and $x_{1}$ via $0=\partial H(r_{+})/\partial y^{b}$. (2) Using these expressions for $r_{+}$ and its derivatives, we compute the necessary derivatives of $S$ with respect to $a$ and $x_{1}$. (3) We then treat $a$ and $x_{1}$ as functions of $M$ and $J$, and implicitly differentiate.  Combining each of these steps, we construct the Hessian and then determine its eigenvalues $h_{i}$ over the range of parameter values where we see the onset of supperradiance.  In Fig.~\ref{fig:Eigen} we display the eigenvalues for the case of the quantum-corrected black hole ($\kappa=-1$).

\begin{center}
\textbf{Method for computing quasinormal mode spectrum}
\end{center}

 \noindent Here we briefly outline how we compute the quasinormal mode (QNM) spectrum for the rotating qBTZ black hole (more detail will be included in a companion article). The essential steps follow closely to those in~\cite{Cartwright:2024iwc}.

 We begin with the naive brane metric (\ref{eq:tbarphibar}) and proceed to remove the conical singularity by imposing bulk regularity conditions, and bring the line element to the canonical form displayed in Eq.~\ref{eq:qbtzrotatmet}. The resulting line element, however, has non-vanishing rotation at the outer horizon, which is inconvenient for infalling Eddington-Finkelstein coordinates. We thus move to co-rotating coordinates by performing the transformation $\bar{\phi}\rightarrow \bar{\phi}+\Omega_+ \bar{t}$ leaving us with
\begin{equation}
    \exd s_{co}^2=g_{\bar{r}\bar{r}}\exd \bar{r}^2+g_{\bar{t}\bar{t}}\exd \bar{t}^2 + 2 g_{\bar{t}\bar{\phi}}\exd \bar{t}\exd \bar{\phi}+g_{\bar{\phi}\bar{\phi}}\exd \bar{\phi}^2\, .
\end{equation}
We then transform to infalling Eddington-Finkelstein coordinates by
\begin{align}
  \exd \bar{t}&=\exd v - \exd \bar{r} \frac{\sqrt{g_{\bar{r}\bar{r}}} \sqrt{g_{\bar{\phi}\bar{\phi}}}}{\sqrt{g_{\bar{t}\bar{\phi}}^2-g_{\bar{t}\bar{t}} g_{\bar{\phi}\bar{\phi}}}} \\
  \exd \bar{\phi}&=\exd \bar{\phi}' + \exd \bar{r} \frac{\sqrt{g_{\bar{r}\bar{r}}} g_{\bar{t}\bar{\phi}}}{\sqrt{g_{\bar{\phi}\bar{\phi}}} \sqrt{g_{\bar{t}\bar{\phi}}^2-g_{\bar{t}\bar{t}} g_{\bar{\phi}\bar{\phi}}}}
\end{align}
Finally, we compactify the radial coordinate $\bar{r}$ by introducing $z=\bar{r}_H/\bar{r}$. The scalar field equations, with a mode decomposition as in Eq.~\ref{eq:mode_expansion} with the appropriate changes, leads to a second order differential equation for $\bar{\Phi}(z)$, which is regular at the horizon as a result of the coordinates, that can be arranged as a generalized eigenvalue problem $M_0\bar{\Phi}=\omega M_1\bar{\Phi}$. Numerical solutions are carried out in Mathematica by discretizing the operators using a pesudo-spectral method comprised of Chebyshev polynomials. An example is shown in Fig.~\ref{fig:QNM}  of the (horizon ingoing) QNM spectrum of a scalar fluctuation with vanishing momentum ($m=0$).

 The QNMs with $\nu=0$ (red dots) agree with the spectrum of classical BTZ~\cite{Cardoso:2001hn,Birmingham:2001hc,Birmingham:2001pj},  i.e.,
\begin{equation}\label{eq:btz_fermion_poles}
  \hspace{-2mm}  \omega= -m -4\pi i T_R(n_z+h_R)\, , \; \omega= m -4\pi i T_L(n_z+h_L)\,,
 \end{equation}
where $n_z\in\mathbb{Z}_+$, $(h_L,h_R)$ are the conformal weights of the dual operator and $T_{L,R}$ are left/right moving temperatures.
The QNMs also display a non-monotonic behavior for small $\nu$ (shown in yellow) before transitioning to monotonic dependence on $\tilde{a}$ as $\nu$ increases (shown at $\nu=0.5$ in purple). An animated GIF of the QNMs as a function of $a$ has been included as ancillary material.

\begin{center}
\textbf{Supplemental displays of superradiant behavior}
\end{center}

\begin{figure*}[t!]
    \centering
    \includegraphics[width=0.49\textwidth]{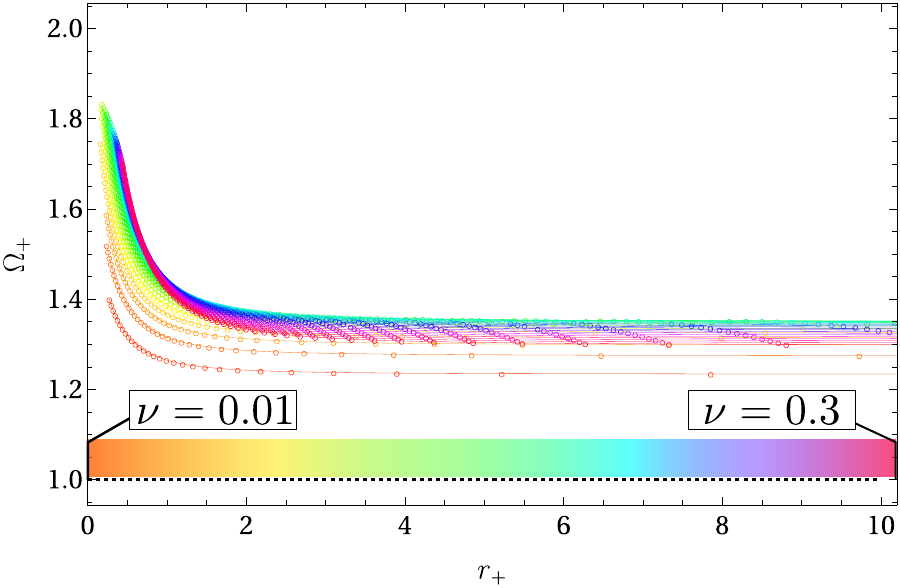}\hfill
    \includegraphics[width=0.495\textwidth]{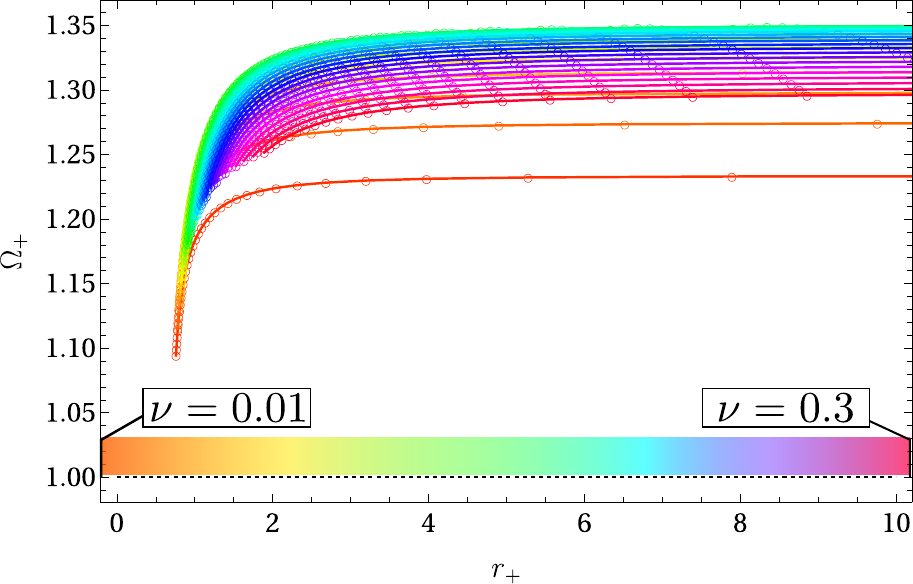 }
    \caption{The onset of superradiance is displayed for the rotating qBTZ black hole. \textit{Left:} $\kappa=+1$. \textit{Right:} $\kappa=-1$. The color of the points represent increasing the strength of the quantum back reaction. The horizontal dashed line displays the threshold $|\Omega_+|\ell_3=1$.
    \label{fig:onset_n_1supp}}
\end{figure*}

\begin{figure}[h]
    \centering
    \includegraphics[scale=0.5]{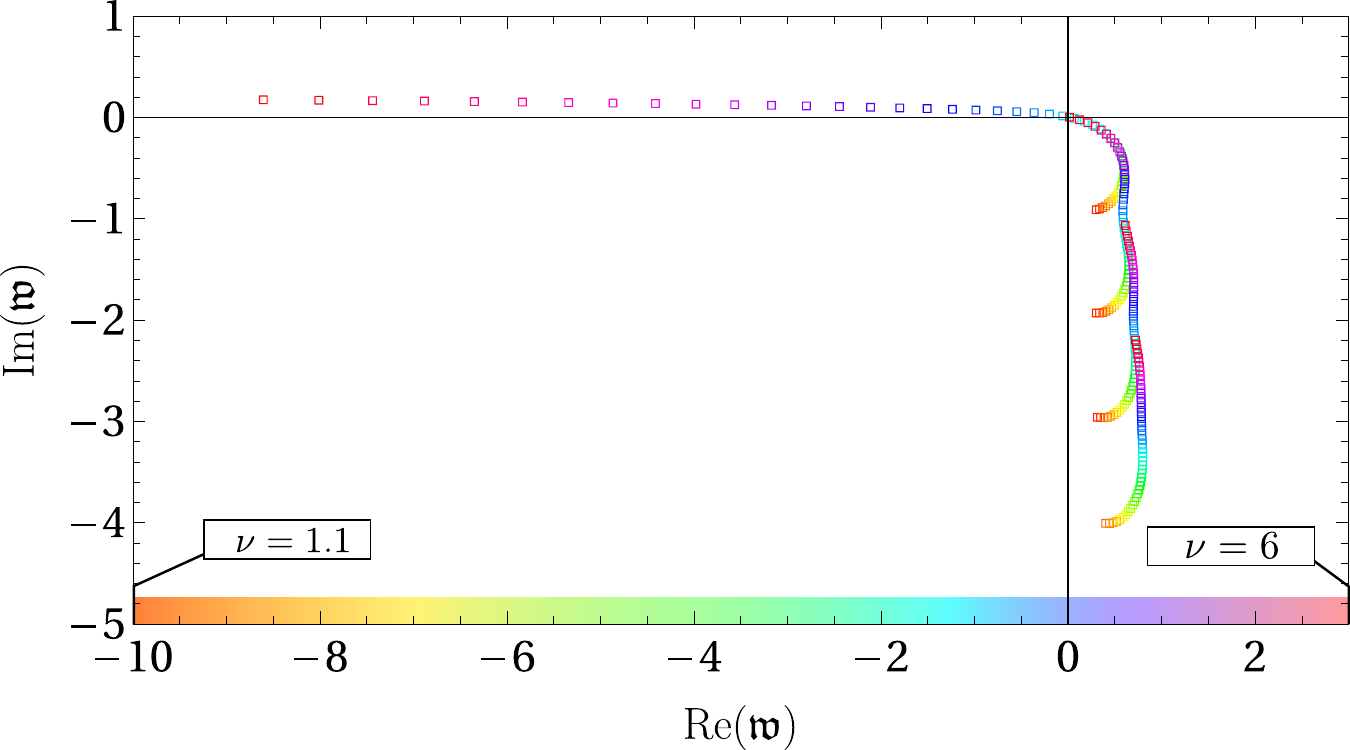}
    \caption{Superradiant QNM frequencies $\mathfrak{w}=\omega/(2\pi T)$ are displayed for the rotating qBTZ black hole with $\kappa=1$ and $m=1$. The color of the points represent increasing the strength of the quantum back reaction with the values of $a$ and $x_1$ dictated by the analytic solution obtained in the supplemental material.
\label{fig:superradiant_modes_analytic_curvesupp}}
\end{figure}

In the main text we displayed the onset of superradiant instability in a plot of $\Omega_{+}$ as a function of $\nu$. It is also illustrative to display the onset via a plot of $\Omega_{+}$ as a function of radius $r_{+}$ over a range $\nu$. We present this in Figure \ref{fig:onset_n_1supp} for both rotating quantum dressed cones ($\kappa=1$) and quantum corrected black holes ($\kappa=-1$). Further, in the main text we presented a plot of superradiant QNM frequencies for both the quantum-dressed cones and quantum-corrected black holes in the limit of weak backreaction. In the case of quantum-dressed cones, we also uncover superradiant QNMs for $\nu>1$, i.e., large (non-perturbative) backreaction -- see Figure \ref{fig:superradiant_modes_analytic_curvesupp}. We do not see such behavior for the quantum-corrected black holes, at least for the lightest modes we consider here.

\end{document}